\documentclass[showpacs,preprintnumbers,amsmath,amssymb,12pt]{revtex4}

\usepackage{amsmath, amssymb, mathrsfs, bbm}
\usepackage{graphicx}
\usepackage{dcolumn}
\usepackage{bm}

\usepackage{bbm}

\newcommand{\beqs}{\begin{equation*}}
\def\beq{\begin{equation}}

\newcommand{\eeqs}{\end{equation*}}
\def\eeq{\end{equation}}

\newcommand{\beqas}{\begin{eqnarray*}}
\newcommand{\beqa}{\begin{eqnarray}}

\newcommand{\eeqas}{\end{eqnarray*}}
\newcommand{\eeqa}{\end{eqnarray}}

\newcommand{\al}{\alpha}

\newcommand{\de}{\delta}

\newcommand{\si}{\sigma}

\newcommand{\Om}{\Omega}

\newcommand{\Si}{\Sigma}

\newcommand{\blist}{\begin{itemize}}

\newcommand{\elist}{\end{itemize}}

\providecommand{\href}[2]{#2}

\DeclareFontFamily{OT1}{rsfs}{}
\DeclareFontShape{OT1}{rsfs}{m}{n}{ <-7> rsfs5 <7-10> rsfs7 <10->rsfs10}{} 
\DeclareMathAlphabet{\mycal}{OT1}{rsfs}{m}{n}

\def\cM{{\cal M}}

\def\TrL2{{\rm Tr}_{L^2}}

\def\atbdry{\Big|_{\partial \cM}}
\def\atbdry0{\Big|_{\partial \cM_0}}
\def\atbdry1{\Big|_{\partial \cM_1}}

\newcommand{\ket}[1]{| #1 \rangle}

\begin{document}


\title{Massive Vertex Operators and Conformal Algebra of\\ the Bosonic Open String Theory in Flat Space-time}


\author{Chuan-Tsung Chan}
\email{ctchan@thu.edu.tw}
\author{Wei-Ming Chen}
\email{tainist@gmail.com}
\affiliation{Department of Physics, Tunghai University, Taiwan, 40704}


\date{\today}

\begin{abstract}
A systematic scheme is developed for solving conformal algebra of the massive vertex operators in the old covariant first quantized string theory. Using the first massive level in the covariant spectrum of bosonic open string theory in flat space-time as an example, we explicitly verify the conformal algebra for the normal-ordered covariant vertex operator. We emphasize that the requirement of fully covariant gauge-fixing specifies the form of normal-ordering for massive vertex operators.
\end{abstract}


\maketitle
\clearpage
\section{Introduction}
String theory in general space-time background typically has a spectrum consisting of infinitely many particles with arbitrary spins. While it is a standard wisdom \cite{Green:1987sp,Lust:1989tj,Scherk:1974jj,Yoneya:1974jg} that in the low-energy approximation ($\al^\prime E^2\rightarrow 0$) string theory reproduces supergravity theory (with stringy corrections), a complete understanding of the quantum dynamics of string theory can not be achieved without the inclusion of massive stringy excitations. This is particularly important if we wish to address the question of a non-perturbative formulation or what is the underlying symmetry principle of string theory \cite{Gross:1988ue,Witten:1988sy}. To study the quantum dynamics of string theory, one often employ the Polyakov world-sheet description and perform the calculations of correlation functions among stringy excitations in the functional integral approach \cite{Polyakov:1981rd,Polyakov:1981re,Hatfield:1992rz}. However, as in the case of standard quantum field theories, the operator formalism sometimes also provides useful insights and different perspective when there exists ambiguity in the functional integral approach. In addition, the Fock space of stringy excitations forms representations of conformal algebra which is of interest to both physicists and mathematicians.\par
In this paper, we address ourselves to the construction of massive vertex operators in the simplest case, namely, bosonic open string theory in 26 dimensional space-time. This old subject has been studied \cite{Weinberg:1985tv,Sasaki:1985py} since early days of string theory, and our original motivation is to apply similar method developed here to the study of string theory in the linear dilaton background \cite{Ho:2007ar,Chan:2009kb}. However, in view of the technical nature, also for pedagogical reason, we believe that it is useful to present our detailed derivations in a separate note. Hopefully, this will help some of the interested readers to really work out the complicated algebra, which is often streamlined in the standard textbooks.\par
We emphasize that, in this paper, we work with fully covariant version of the operator formalism. Here, the only gauge fixing is the world-sheet conformal gauge, and the requirement of conformal invariance (Virasoro constraints) makes the normal-ordering of massive vertex operators a subtle issue. To the best of our knowledge, this issue has not been discussed in the literature explicitly. We hope that our presentation helps in clarifying this non-trivial issue.\par
This paper is organized as follows. In section \ref{II}, we review some of the basic notations in the bosonic open string theory in flat space-time, together with a discussion of physical state conditions and covariant spectrum. Conformal algebra of tachyon and photon are presented in sections \ref{III} and \ref{IV}, respectively. In section \ref{V}, we study the conformal algebra for the first massive vertex operator in bosonic open string theory. We summarize our result in section \ref{VI}. Some of the useful equations are collected in Appendix \ref{A}.
\section{Old Covariant Quantization of Bosonic Open String Theory in Flat Space-time}\label{II}
\subsection{Basic Notations}
The Polyakov action for the bosonic open string theory in flat space-time is given by
\beqa
S[X^\mu, h]&=&\frac{1}{4\pi\al^\prime}\int_\Si d^2 \si \sqrt{h}h^{ab} \partial_aX(\si)\cdot\partial_bX(\si).
\eeqa
From the action we can extract the energy-momentum tensor ($z=e^{-i\si^1+\si^2}$, $\si^2=i\tau$),
\beqa
T_{zz}=-\frac{1}{\al^\prime}:\partial X\cdot\partial X:.
\eeqa
The string coordinates in the oscillator representation are
\beqa
X^\mu(z,\bar{z})&=&x^\mu-i\al^\prime p^\mu \ln|z|^2+i\sqrt{\frac{\al^\prime}{2}}\sum_{m=-\infty, m\neq0}^{m=\infty}\frac{\al_m^\mu}{m}(z^{-m}+\bar{z}^{-m}).
\eeqa
The Virasoro generators of the conformal transformation are defined as the Fourier modes of the energy-momentum tensor $T_{zz}$,
\beqa
L_m\equiv\oint dz z^{m+1}T_{zz}=\frac{1}{2}\sum_{n=-\infty}^{\infty}:\al_{m-n}\al_n:.
\eeqa
Using the basic commutation relation among oscillators, $\big[\al_m,\al_n\big]=m\de_{m+n}$, one can check that the Virasoro generators satisfy the following algebraic relation \cite{Green:1987sp,Lust:1989tj},
\beqa
\big[L_m,L_n\big]=\big(m-n\big)L_{m+n}+\frac{D}{12}m\big(m^2-1\big)\de_{m+n}.
\eeqa

In our later calculations of conformal algebra, Eq.\eqref{ComforAlg}, for various normal-ordered vertex operators, it is useful to decompose the string coordinates into the creation and annihiliation parts. For this purpose, we define ($y\equiv e^{i\tau}$)
\beqa
\notag X^\mu&\equiv& X^\mu_-+X^\mu_0+X^\mu_+,\\
\mbox{annihilation part of $X^\mu$}~&\Rightarrow&~ X^\mu_+\equiv\sqrt{2\al^\prime}\sum_{n=1}^\infty\frac{i}{n}\al_{n}e^{-in\tau}=\sqrt{2\al^\prime}\sum_{n=1}^\infty\frac{i}{n}\al_{n}y^{-n},\\
\mbox{zero mode part of $X^\mu$}~&\Rightarrow&~ X^\mu_0\equiv x^\mu+2\al^\prime p^\mu\tau,\\
\mbox{creation part of $X^\mu$}~&\Rightarrow&~ X^\mu_-\equiv-\sqrt{2\al^\prime}\sum_{n=1}^\infty\frac{i}{n}\al_{-n}e^{in\tau}=-\sqrt{2\al^\prime}\sum_{n=1}^\infty\frac{i}{n}\al_{n}y^n.
\eeqa
It is useful to define the combinations of $X^\mu_+$, $X^\mu_0$, $X^\mu_-$, and we have
\beqa
Y^\mu&\equiv&X^\mu_-+X^\mu_0,\\
Z^\mu&\equiv&X^\mu_0+X^\mu_+.
\eeqa
The commutation relations for oscillators and the string coordinates are given by
\beqa
\begin{array}{ll}
\big[\al_n^\mu,X^\nu_-(\tau)\big]=-i\sqrt{2\al^\prime}y^n\eta^{\mu\nu}&,~~~~~n>0,\\
\big[\al_0^\mu,X^\nu_0(\tau)\big]=-i\sqrt{2\al^\prime}\eta^{\mu\nu}&,~~~~~n=0,\\
\big[\al_n^\mu,X^\nu_+(\tau)\big]=-i\sqrt{2\al^\prime}y^n\eta^{\mu\nu}&,~~~~~n<0.
\end{array}
\eeqa
More generally,
\beqa
\label{alXZ}\big[\al_n^\mu,X^\nu(\tau)\big]=-i\sqrt{2\al^\prime}y^n\eta^{\mu\nu}~~,~~\forall n\in Z .
\eeqa
The commutation relations for the Virasoro generators and the string coordinates are (in the following, we always assume $m>0$.)
\beqa
\notag\big[L_m,X^\mu_+(\tau)\big]&=&-i\sqrt{2\al^\prime}\sum_{n=m+1}^{\infty}y^{m-n}\al_n^\mu,\\
\big[L_m,X^\mu_0(\tau)\big]&=&-i\sqrt{2\al^\prime}\al_m^\mu,\\
\notag\big[L_m,X^\mu_-(\tau)\big]&=&-i\sqrt{2\al^\prime}\sum_{n=-\infty}^{m-1}y^{m-n}\al_n^\mu.
\eeqa
More generally,
\beqa
\big[L_m,X^\mu(\tau)\big]&=&-iy^m\dot X^\mu,\\
\big[L_m,\dot X^\mu(\tau)\big]&=&my^m\dot X^\mu-iy^m\ddot X^\mu,\\
\big[L_m,\ddot X^\mu(\tau)\big]&=&im^2y^m\dot X^\mu+2my^m\ddot X^\mu-iy^m\dddot X^\mu.
\eeqa
We collect other useful formulae, which can be derived from the results above in Appendix \ref{A}.
\subsection{Covariant Spectrum of Bosonic Open String Theory in Flat Space-time}
To calculate the physical state spectrum of the bosonic open string theory in the flat space-time, we solve all possible linear combinations of creation operators acting on a Fock vacuum, subject to the Virasoro constraints:
\beqa
L_0\ket{\Phi(k)}=\ket{\Phi(k)},\quad \mbox{and} \quad L_n \ket{\Phi(k)}=0,\quad n\geqslant1.
\eeqa
These constraints in general lead to the generalized on-shell condition for the center of mass momenta, and restrict the polarization tensors to be transverse and traceless.
In the covariant spectrum, where physical states consist of various linear combinations of oscillators with polarization tensors, Virasoro constraints implies lower spin polarization tensors are given by the projections of higher spin polarization tensors \cite{Lee:1994wp,Chan:2005qd}.
In this paper, we shall focus on the physical states up to the first massive level, and we shall use capital letters to represent the particles. For instance, the tachyon state (T) is defined as
\beqa
\ket{T(k)}\equiv\ket{0,k}.
\eeqa
The $L_0$ condition
\beqa
\label{L0T}L_0\ket{T(k)}=\frac{1}{2}\al_0^2\ket{0,k}=\ket{T(k)},
\eeqa
together with the eigenvalue condition for $\al_0^\mu$, $\al_0^\mu\ket{k,0}=\sqrt{2\al^\prime}k^\mu\ket{k,0}$, leads to generalized on-shell condition
\beqa
\label{OnST}\al^\prime k^2=1.
\eeqa
The $L_1$ and $L_2$ conditions are trivial for the tachyon state.
\par
At massless level, we have a photon state (P) with polarization vector $\zeta(k)$,
\beqa
\ket{P(\zeta, k)}\equiv\zeta\cdot\al_{-1}\ket{0,k}.
\eeqa
One can check that the $L_0$ condition,
\beqa
L_0\ket{P(\zeta, k)}=(\al_{-1}\cdot\al_{1})\zeta\cdot\al_{-1}\ket{0,k}=\ket{P(\zeta, k)},
\eeqa
leads to
\beqa
\label{L0P}\al^\prime k^2=0.
\eeqa
On the other hand, the $L_1$ condition
\beqa
L_1\ket{P(\zeta, k)}=(\al_{0}\cdot\al_{1})\zeta\cdot\al_{-1}\ket{0,k}=0,
\eeqa
gives the generalized transverse condition,
\beqa
\label{L1P}\zeta\cdot k=0.
\eeqa
The $L_2$ condition is trivial for the photon state.
\par Finally, at the first massive level (M), we have a tensor particle with spin-two, and it is written as
\beqa
\ket{M(\epsilon_{\mu\nu},k)}\equiv\big(\epsilon_{\mu\nu}\al^\mu_{-1}\al^\nu_{-1}+\epsilon_{\mu}\al^\mu_{-2}\big)\ket{0,k}.
\eeqa
The generalized on-shell condition, as derived from the $L_0$ condition
\beqa
L_0\ket{M(\epsilon_{\mu\nu},k)}=(\al_{-1}\cdot\al_{1}+\al_{-2}\cdot\al_{2})\Big(\epsilon_{\mu\nu}\al^\mu_{-1}\al^\nu_{-1}+\epsilon_{\mu}\al_{-2}^\mu\Big)\ket{0,k}=\ket{M(\epsilon_{\mu\nu},k)},
\eeqa
is
\beqa
\label{L0M}\al^\prime k^2=-1.
\eeqa
The $L_{1}$ conditions,
\beqa
L_{1}\ket{M(\epsilon_{\mu\nu},k)}=
(\al_{0}\cdot\al_{1}+\al_{-1}\cdot\al_{2})\Big(\epsilon_{\mu\nu}\al^\mu_{-1}\al^\nu_{-1}+\epsilon_{\mu}\al_{-2}^\mu\Big)\ket{0,k}=0,
\eeqa
implies that the polarization vector $\epsilon_\mu$ can be written as a projection of the spin-two polarization tensor $\epsilon_{\mu\nu}$,
\beqa
\label{L1M}\sqrt{2\al^\prime}\epsilon_{\mu\nu}k^\nu+\epsilon_\mu=0.
\eeqa
For this reason, we suppress the $\epsilon_\mu$ dependence in the notation of $M(\epsilon_{\mu\nu},k)$.
Finally, the $L_2$ conditions, $L_{2}\ket{M(\epsilon_{\mu\nu},k)}=0$,
\beqa
\Big(\al_{1}\cdot\al_{1}+\al_{0}\cdot\al_{2}\Big)\Big(\epsilon_{\mu\nu}\al^\mu_{-1}\al^\nu_{-1}+\epsilon_{\mu}\al_{-2}^\mu\Big)\ket{0,k}=0,
\eeqa
gives
\beqa
\label{L2M}\epsilon_{\mu\nu}\eta^{\mu\nu}+2\sqrt{2\al^\prime}\epsilon_\mu k^\mu=0.
\eeqa
Substituting $\epsilon_\mu$ from Eq.\eqref{L1M} to Eq.\eqref{L2M}, we get
\beqa
\label{Vra}4\al^\prime\epsilon_{\mu\nu}k^\mu k^\nu-\epsilon_{\mu\nu}\eta^{\mu\nu}=0.
\eeqa
\par
The main point of this paper is to show that, by choosing a suitable normal-ordered vertex operator for each stringy state, we can derive the same physical state conditions, Eqs.\eqref{OnST}, \eqref{L0P}, \eqref{L1P}, \eqref{L0M}, \eqref{L1M}, \eqref{L2M} through conformal algebra Eq.\eqref{ComforAlg}.
\subsection{Vertex Operators and Conformal Algebra}
Our main goal in this paper is to solve for the covariant vertex operators of bosonic open string theory. To construct general vertex operators, it is useful to recall the correspondence between states and operators in conformal field theory. In the oscillator representation for the one string Fock space, we have
\beqas
\al_{m}^\mu&=&\Big(\frac{2}{\al^\prime}\Big)^{\frac{1}{2}}\oint\frac{dz}{2\pi}z^{-m}\partial^m X^\mu(z)\\
&\rightarrow& \Big(\frac{2}{\al^\prime}\Big)^{\frac{1}{2}}\frac{i}{(m-1)!}\partial^mX^\mu(0),\\
x_0^\mu&\rightarrow&X^\mu(0).
\eeqas
From this correspondence, we have the following dictionary for bosonic open string states and the correspondence vertex operators:
\beqas
\begin{array}{rrcl}
\mbox{tachyon (T)}:&\ket{0,k}&\sim&:e^{ik\cdot X}:\\
\mbox{photon (P)}:&\zeta\cdot\al_{-1}\ket{k,0}&\sim&\zeta\cdot \partial X :e^{ik\cdot X}:\\
\mbox{first massive state (M)}:\quad&\big(\epsilon_{\mu\nu}\al^\mu_{-1}\al^\nu_{-1}+\epsilon_{\mu}\al_{-2}^{\mu}\big)\ket{k,0}
&\sim&\Big[\epsilon_{\mu\nu}(\partial X^\mu)(\partial X^\nu)+\epsilon_{\mu}\partial^2 X^\mu\Big]:e^{ik\cdot X}:\quad.
\end{array}
\eeqas
Note that the partial derivative $\partial$ denotes the differentiation with respect to complex world-sheet variable in the upper half plane $\partial\equiv\frac{\partial}{\partial z}$. In the strip diagram for open string world-sheet, we need to make a change of variable $z=e^{i\tau}$ to facilitate operator calculations (this also leads to some factors of $i$ in the expression, see Eq.\eqref{VMM}).
\par In the calculations of scattering amplitudes (correlation functions) in any conformal field theory, the use of vertex operators ensures that the final results are conformal invariant. For string theory, in particular, we use integrated vertex operators to allow for all possible particle emissions (or absorptions)
\beqas
V_{string}\equiv\int d\tau V_{string}(\tau).
\eeqas
Here to compensate for the conformal transformation of the integration measure
$d\tau$ $\rightarrow$ $d\tau^\prime$, we need to impose the condition that the unintegrated vertex operator $V_{string}(\tau)$ to transform like $V_{string}(\tau)~\rightarrow~\Big(\frac{d\tau}{d\tau^\prime}\Big)V_{string}(\tau)$. Consequently, we require all unintegrated vertex operators to carry conformal dimension $J=1$. More precisely, if we check the action of conformal transformation induced by energy-momentum tensor on any unintegrated vertex operator, we must have the following algebraic relation \cite{Green:1987sp},
\beqa
\label{ComforAlg}\big[L_m,V_{string}(\tau)\big]=e^{im\tau}\Big(-i\frac{d}{d\tau}+mJ\Big)V_{string}(\tau).
\eeqa
In the following, we show that the solutions of unintegrated vertex operators to the conformal algebra, Eq.\eqref{ComforAlg}, must satisfy the Virasoro constraints.
\section{Conformal Algebra for the Tachyon (T) Vertex Operator}\label{III}
The normal-ordered vertex operator for tachyon (T) is defined as
\beqa
\label{VT}V_T\equiv:e^{ik\cdot X}:= e^{ik\cdot X_-}e^{ik\cdot Z}.
\eeqa
Henceforth, to simplify the calculations, we have combined the zero mode part $X_0$ with annihilation operator $X_+$, $Z\equiv X_0+X_+$.\par
Using the formulae in Appendix \ref{A}, Eqs.\eqref{A3}.\eqref{A4}, we can compute the conformal algebra of the vertex operator for tachyon,
\beqa
\label{LVT}\big[L_m,V_T\big]=e^{ik\cdot X_-}\big[L_m,e^{ik\cdot Z}\big]+\big[L_m,e^{ik\cdot X_-}\big]e^{ik\cdot Z}
=y^me^{ik\cdot X_-}Ue^{ik\cdot Z}.
\eeqa
Here
\beqa
\notag U&\equiv&\sqrt{2\al^\prime}\sum_{n=m}^{\infty}(k\cdot\al_n)y^{-n}+\sqrt{2\al^\prime}\sum_{n=1}^{m-1}(k\cdot\al_n)y^{-n}+k\cdot \dot Y+\al^\prime k^2(m-1)\\
\notag &=&k\cdot \dot X_++k\cdot \dot Y-\al^\prime k^2+m(\al^\prime k^2)\\
\label{U}&=&k\cdot\dot X_-+k\cdot\dot Z-\al^\prime k^2+m(\al^\prime k^2).
\eeqa
Substituting this result into commutator Eq.\eqref{LVT}. we get
\beqa
\notag&&\big[L_m,V_T\big]\\
\notag&=&y^m\Big[e^{ik\cdot X_-}(k\cdot\dot X_-)e^{ik\cdot Z}+e^{ik\cdot X_-}(k\cdot\dot Z-\al^\prime k^2)e^{ik\cdot Z}+m\al^\prime k^2V_T\Big]\\
\notag&=&y^m\Big[\big(-i\frac{d}{d\tau}e^{ik\cdot X_-}\big)e^{ik\cdot Z}+e^{ik\cdot X_-}\big(-i\frac{d}{d\tau}e^{ik\cdot X_0}\big)e^{ik\cdot X_+}+e^{ik\cdot Y}\big(-i\frac{d}{d\tau}e^{ik\cdot X_+}\big)+m\al^\prime k^2 V_T\Big]\\
&=&y^m\Big(-i\frac{d}{d\tau}+m\al^\prime k^2\Big)V_T.
\eeqa
From this result, it is clear that if we require the unintegrated vertex operator $V_T$ to have conformal dimension $J=1$, we recover the on-shell condition for the tachyon state, $\al^\prime k^2=1$, Eq.\eqref{OnST}.
\section{Conformal Algebra for the Photon (P) Vertex Operator}\label{IV}
The normal-ordered vertex operator for photon (P) is defined as

\beqa
\notag V_P&\equiv& \frac{\zeta\cdot\dot{X}}{\sqrt{2\al^\prime}}V_T\\
\notag&=&\frac{\zeta_{\mu}}{\sqrt{2\al^\prime}}\big(\dot{X}^\mu_-V_T+\dot{X}^\mu_0V_T+V_T\dot{X}^\mu_+\big)+
\label{VP}\frac{\zeta_\mu}{\sqrt{2\al^\prime}}\big[\dot X_+^\mu, V_T\big]\\
&=&\frac{\zeta_\mu}{\sqrt{2\al^\prime}} V_{P_1}^\mu-\sqrt\frac{\al^\prime}{2}(\zeta\cdot k)V_T.
\eeqa
Here we define
\beqa
V_{P_1}^\mu&\equiv&:\frac{\dot{X}^\mu}{\sqrt{2\al^\prime}}V_T:=\dot{X}_-^\mu V_T+\dot{X}_0^\mu V_T+V_T\dot{X}_+^\mu=e^{ik\cdot X_-}\dot X^\mu e^{ik\cdot Z}.
\eeqa
The commutator $\big[\dot X^\mu_+, V_T\big]=-\al^\prime k^\mu V_T$ in Eq.\eqref{VP} can be derived using Eq.\eqref{A-1} from Appendix \ref{A}. Here we observe a general pattern, that is, the normal-ordered vertex operator $V_{P}$ is not equal to the naive guess, $\frac{\zeta_\mu}{\sqrt{2\al^\prime}}V_{P_1}^\mu$. There are contributions of vertex operators with lower spin. For this reason, we call Eq.\eqref{VP} the first descending formula.\par Using the formulae in Appendix \ref{A}, Eqs.\eqref{A3},\eqref{A4}, we can compute the conformal algebra of the vertex operator for photon,
\beqa
\notag\big[L_m,V_{P_1}^\mu\big]&=&e^{ik\cdot X_-}\dot X^\mu\big[L_m,e^{ik\cdot Z}\big]+\big[L_m,e^{ik\cdot X_-}\big]\dot X^\mu e^{ik\cdot Z}+e^{ik\cdot X_-}\big[L_m,\dot X^\mu\big]e^{ik\cdot Z}\\
\label{LVPP}&=&y^me^{ik\cdot X_-}U_1^\mu e^{ik\cdot Z},
\eeqa
where
\beqas
U_1^\mu&\equiv&\dot X^\mu\Big[\sqrt{2\al^\prime}\sum_{n=m}^{\infty}(k\cdot\al_n)y^{-n}\Big]+\Big[\sqrt{2\al^\prime}\sum_{n=1}^{m-1}(k\cdot\al_n)y^{-n}\Big]\dot X^\mu+(k\cdot \dot Y-\al^\prime k^2)\dot X^\mu\\
&&+m\al^\prime k^2\dot X^\mu+m\dot X^\mu-i\ddot X^\mu.
\eeqas
Computing the commutator for $k\cdot \al_n$ and $\dot X^\mu$ using Eq.\eqref{alXZ}, we have
\beqa
\label{U1}U_1^\mu=\dot X^\mu(k\cdot\dot X_+)+(k\cdot \dot Y-\al^\prime k^2)\dot X^\mu-i\ddot X^\mu+m(\al^\prime k^2+1)\dot X^\mu+\al^\prime(m-1)mk^\mu.
\eeqa
Substituting Eq.\eqref{U1} into the conformal algebra for $V_{P_1}^\mu$, Eq.\eqref{LVPP}, we get
\beqa
\notag\big[L_m,V_{P_1}^\mu\big]&=&y^m\left[\begin{array}{rl}&e^{ik\cdot X_-}\dot X^\mu(k\cdot\dot X_+)e^{ik\cdot Z}+e^{ik\cdot X_-}(k\cdot\dot Y-\al^\prime k^2)\dot X^\mu e^{ik\cdot Z}\\
+&e^{ik\cdot X_-}(-i\ddot X^\mu) e^{ik\cdot Z}
+m(\al^\prime k^2+1)V_{P_1}^\mu+\al^\prime(m^2-m)k^\mu V_T\end{array}\right]\\
\label{LVP1}&=&y^m\bigg\{\Big[-i\frac{d}{d\tau}+m(\al^\prime k^2+1)\Big]V_{P_1}^\mu+\al^\prime(m^2-m)k^\mu V_T\bigg\}
\eeqa
If we add the tachyon contribution in Eq.\eqref{VP} and the $V^\mu_{P_1}$ contribution from Eq.\eqref{LVP1} to obtain the full conformal algebra for $V_P$, we get
\beqa
\notag\big[L_m,V_P\big]&=&\frac{\zeta_\mu}{\sqrt{2\al^\prime}}\big[L_m,V_{P_1}^\mu\big]-\sqrt\frac{\al^\prime}{2}(\zeta\cdot k)\big[L_m,V_T\big]\\
\label{LMMVP}&=&y^m\bigg\{\Big[-i\frac{d}{d\tau}+m(\al^\prime k^2+1)\Big]V_{P}+\al^\prime m^2(\zeta\cdot k) V_T\bigg\}.
\eeqa
From this result, we naturally derive the ($L_0$) on-shell condition, $\al^\prime k^2=0$, and the ($L_1$) transverse condition, $\zeta\cdot k=0$.
\section{Conformal Algebra for the First Massive Spin-two (M) Vertex Operator}\label{V}
\subsection{Definition and Descending Formula for the Vertex Operator}
The normal-ordered vertex operator corresponding to the first massive spin-two state (M) is defined as
\beqa
\notag V_M&\equiv&\frac{\epsilon_{\mu\nu}}{2\al^\prime}\dot X^\mu\dot X^\nu V_T-\frac{i\epsilon_\mu}{\sqrt{2\al^\prime}} \ddot{X}^\mu V_T\\
\notag &=&\frac{\epsilon_{\mu\nu}}{2\al^\prime}\left(\begin{array}{rl}&\dot{X}_-^\mu\dot{X}_-^\nu V_T
+\dot{X}_-^\mu\dot{X}_0^\nu V_T+\dot{X}_0^\mu\dot{X}_-^\nu V_T\\
+&\dot{X}_0^\mu\dot{X}_0^\nu V_T+\dot{X}_0^\mu V_T\dot{X}_+^\nu
+\dot{X}_0^\nu V_T\dot{X}_+^\mu\\
+&{X}_-^\mu V_T \dot{X}_+^\nu
+\dot{X}_-^\nu V_T\dot{X}_+^\mu+ V_T \dot{X}_+^\mu\dot{X}_+^\nu
\end{array}\right)-\frac{i\epsilon_\mu}{\sqrt{2\al^\prime}}\big(\ddot{X}^\mu_-V_T+V_T\ddot{X}^\mu_+\big)\\
\label{VMM}&&+\frac{\epsilon_{\mu\nu}}{2\al^\prime}\left\{\begin{array}{rl}&\big[\dot X_+^\mu \dot X_+^\nu, V_T\big]+\big[\dot X_+^\mu,\big(\dot X_0^\nu+\dot X_-^\nu\big)V_T\big]\\
+&\big(\dot X_0^\nu+\dot X_-^\nu\big)\big[\dot X_+^\nu,V_T\big]\end{array}\right\}-\frac{i\epsilon_\mu}{\sqrt{2\al^\prime}}\big[\ddot X_+^\mu,V_T\big].
\eeqa
The contribution from the commutator terms in Eq.\eqref{VMM} can be calculated using the formulae, Eqs.\eqref{A-1}, \eqref{A-2}, \eqref{A-3}, in Appendix \ref{A}. The results are
\beqas
\big[\dot X_+^\mu \dot X_+^\nu, V_T\big]&=&-\al^\prime k^\mu V_TX_+^\nu-\al^\prime k^\nu V_T X_+^\mu+\al^\prime k^\mu k^\nu V_T,\\
\big[\dot X_+^\mu,\big(\dot X_0^\nu+\dot X_-^\nu\big)V_T\big]&=&-\al^\prime k^\mu \big(\dot X_0^\nu+\dot X_-^\nu\big)V_T-\frac{\al^\prime}{6}\eta^{\mu\nu}V_T,\\
\big(\dot X_0^\nu+\dot X_-^\nu\big)\big[\dot X_+^\nu,V_T\big]&=&-\al^\prime k^\nu\big(\dot X_0^\mu+\dot X_-^\mu\big)V_T.
\eeqas
After simplify the commutator terms, we get the second descending formula for the normal-ordered vertex operator
\beqa
\label{VM}V_M=\tilde V_{M}-\epsilon_{\mu\nu}k^\mu V_{P_1}^\nu+\Om V_T,
\eeqa
where
\beqa
\label{Om}\Om\equiv\frac{\al^\prime}{2}\epsilon_{\mu\nu}k^\mu k^\nu
-\frac{1}{12}\epsilon_{\mu\nu}\eta^{\mu\nu}+\frac{\sqrt{2\al^\prime}\epsilon\cdot k }{12},
\eeqa
and
\beqa
\notag\hspace{-0.5cm}\tilde V_{M}(\epsilon_{\mu\nu},\epsilon_\mu)&\equiv&\frac{\epsilon_{\mu\nu}}{2\al^\prime}\left(\begin{array}{rl}&\dot{X}_-^\mu\dot{X}_-^\nu V_T
+\dot{X}_-^\mu\dot{X}_0^\nu V_T+\dot{X}_0^\mu\dot{X}_-^\nu V_T\\
+&\dot{X}_0^\mu\dot{X}_0^\nu V_T+\dot{X}_0^\mu V_T\dot{X}_+^\nu
+\dot{X}_0^\nu V_T\dot{X}_+^\mu\\
+&{X}_-^\mu V_T \dot{X}_+^\nu
+\dot{X}_-^\nu V_T\dot{X}_+^\mu+ V_T \dot{X}_+^\mu\dot{X}_+^\nu
\end{array}\right)-\frac{i\epsilon_\mu}{\sqrt{2\al^\prime}}(\ddot{X}^\mu_-V_T+V_T\ddot{X}^\mu_+\big)\\
\label{TVM}&=&\frac{\epsilon_{\mu\nu}}{2\al^\prime}V_{M_1}^{\mu\nu}-i\frac{\epsilon_\mu}{\sqrt{2\al^\prime}}V_{M_1}^\mu,
\eeqa
where
\beqa
\notag V_{M_1}^{\mu\nu}&\equiv&:\dot X^\mu\dot X^\nu V_T:\\
\notag&=&e^{ik\cdot X_-}\dot Y^\mu\dot Y^\nu e^{ik\cdot Z}+e^{ik\cdot X_-}\dot Y^\mu\dot X^\nu_+ e^{ik\cdot Z}\\
\notag &&+e^{ik\cdot X_-}\dot Y^\nu\dot X^\mu_+ e^{ik\cdot Z}+e^{ik\cdot X_-}\dot X^\mu_+\dot X^\nu_+ e^{ik\cdot Z}\\
&=&e^{ik\cdot X_-}W^{\mu\nu} e^{ik\cdot Z},\\
V_{M_1}^\mu&\equiv&:\ddot{X}^\mu V_T:=e^{ik\cdot X_-}\ddot X^\mu e^{ik\cdot Z},\\
W^{\mu\nu}&\equiv&\dot Y^\mu\dot X^\nu+\dot X^\nu\dot X^\mu_+.
\eeqa
\subsection{Computation of the Conformal Algebra $\big[L_m,V_{M_1}^{\mu\nu}\big]$}
From the definition of the vertex operator $V_{M_1}^{\mu\nu}\equiv e^{ik\cdot X_-}W^{\mu\nu} e^{ik\cdot Z}$, and the formulae, Eqs.\eqref{A3}, \eqref{A4}, \eqref{A5}, in Appendix \ref{A}, we can calculate the conformal algebra for $V_{M_1}^{\mu\nu}$.
\beqa
\notag\big[L_m,V_{M_1}^{\mu\nu}\big]&=&e^{ik\cdot X_-}W^{\mu\nu}\big[L_m,e^{ik\cdot Z}\big]+\big[L_m,e^{ik\cdot X_-}\big]W^{\mu\nu} e^{ik\cdot Z}+e^{ik\cdot X_-}\big[L_m,W^{\mu\nu}\big]e^{ik\cdot Z}\\
\label{VM1}&=&y^me^{ik\cdot X_-}U^{\mu\nu} e^{ik\cdot Z},
\eeqa
where
\beqas
U^{\mu\nu}&\equiv& W^{\mu\nu}\Big[\sqrt{2\al^\prime}\sum_{n=m}^{\infty}(k\cdot\al_n)y^{-n}\Big]+\Big[\sqrt{2\al^\prime}\sum_{n=1}^{m-1}(k\cdot\al_n)y^{-n}\Big]W^{\mu\nu}+(k\cdot \dot Y)W^{\mu\nu}\\
&&+\al^\prime k^2(m-1)W^{\mu\nu}-i\frac{d}{d\tau}W^{\mu\nu}+2mW^{\mu\nu}+\frac{\al^\prime }{3}m(m^2-1)\eta^{\mu\nu}.
\eeqas
Computing the commutator for $k\cdot \al_n$ and $W^{\mu\nu}$, Eq.\eqref{A6}, we get
\beqa
\notag U^{\mu\nu}&=&(k\cdot \dot Y)W^{\mu\nu}-i\frac{d}{d\tau}W^{\mu\nu}+W^{\mu\nu}(k\cdot\dot X_+)+\Big[m(\al^\prime k^2+2)-\al^\prime k^2\Big]W^{\mu\nu}\\
\label{Umunu}&&+\al^\prime(m^2-m)\big(k^\mu\dot X^\nu+k^\nu\dot X^\mu\big)+\frac{\al^\prime}{3}(m^3-m)\eta^{\mu\nu}.
\eeqa
Substituting the result for $U^{\mu\nu}$ into Eq.\eqref{VM1}, we have
\beqa
\notag\big[L_m,V_{M_1}^{\mu\nu}\big]&=&y^m
\left[\begin{array}{rl}
&e^{ik\cdot X_-}(k\cdot\dot Y-\al^\prime k^2)W^{\mu\nu}e^{ik\cdot Z}+e^{ik\cdot X_-}\Big(\displaystyle-i\frac{d}{d\tau}W^{\mu\nu}\Big)e^{ik\cdot Z}\\+&e^{ik\cdot X_-}W^{\mu\nu}(k\cdot \dot X_+)e^{ik\cdot Z}
+m(\al^\prime k^2+2)V_{M_1}^{\mu\nu}\\+&\al^\prime(m^2-m)\big(k^\mu V_{P_1}^\nu+k^\nu V_{P_1}^\mu\big)+\displaystyle\frac{\al^\prime}{3}(m^3-m)\eta^{\mu\nu}V_T
\end{array}\right]\\
\label{LVM12}&=&y^m\left\{\begin{array}{rl}&\Big[\displaystyle-i\frac{d}{d\tau}+m(\al^\prime k^2+2)\Big]V_{M_1}^{\mu\nu}\\
+&\al^\prime(m^2-m)\big(k^\mu V_{P_1}^\nu+k^\nu V_{P_1}^\mu\big)+\displaystyle\frac{\al^\prime}{3}(m^3-m)\eta^{\mu\nu}V_T\end{array}\right\}.
\eeqa
\subsection{Computation of the Conformal Algebras $\big[L_m,V_{M_1}^{\mu}\big]$ and $\big[L_m,\tilde V_{M}\big]$}
\par Calculations of the conformal algebra associated with the second term in Eq.\eqref{TVM} is very similar to that of photon vertex operator, Eq.\eqref{LVP1},
\beqa
\notag\big[L_m,V_{M_1}^\mu\big]&=&e^{ik\cdot X_-}\ddot X^\mu\big[L_m,e^{ik\cdot Z}\big]+\big[L_m,e^{ik\cdot X_-}\big]\ddot X^\mu e^{ik\cdot Z}+e^{ik\cdot X_-}\big[L_m,\ddot X^\mu\big]e^{ik\cdot Z}\\
\label{LVM11}&=&y^me^{ik\cdot X_-}U_2^\mu e^{ik\cdot Z},
\eeqa
where
\beqas
U_2^\mu&\equiv&\ddot X^\mu\Big[\sqrt{2\al^\prime}\sum_{n=m}^{\infty}(k\cdot\al_n)y^{-n}\Big]+\Big[\sqrt{2\al^\prime}\sum_{n=1}^{m-1}(k\cdot\al_n)y^{-n}\Big]\ddot X^\mu+(k\cdot \dot Y)\ddot X^\mu\\
&&+(m-1)\al^\prime k^2\ddot X^\mu+(-i\dddot X^\mu+2m\ddot X^\mu+im^2\dot X^\mu).
\eeqas
Computing the commutator for $k\cdot\al_n$ and $\ddot X^\mu$ using Eq.\eqref{alXZ}, we get
\beqa
\notag U_2^\mu&=&(k\cdot \dot Y-\al^\prime k^2)\ddot X^\mu-i\dddot X^\mu+\ddot X^\mu(k\cdot\dot X^\mu_+)\\
\label{UU2}&&+m(\al^\prime k^2+2)\ddot X^\mu+im\dot X^\mu+\frac{1}{3}(2m^3-3m^2+m)(i\al^\prime k^\mu).
\eeqa
Substituting the result for $U^\mu_2$ into Eq.\eqref{LVM11}, we have
\beqa
\notag\big[L_m,V_{M_1}^\mu\big]&=&y^m\left[\begin{array}{rl}
                              & e^{ik\cdot X_-}(k\cdot \dot Y-\al^\prime k^2)\ddot X^\mu e^{ik\cdot Z}+e^{ik\cdot X_-}(-i\dddot X^\mu)\ddot X^\mu e^{ik\cdot Z} \\
                              + & e^{ik\cdot X_-}\ddot X^\mu(k\cdot\dot X^\mu_+)\ddot X^\mu e^{ik\cdot Z}+m(\al^\prime k^2+2) V_{M_1}^\mu\\
                              + & im^2V^\mu_{P_1}+\displaystyle \frac{1}{3}(2m^3-3m^2+m)i\al^\prime k^\mu V_T
                            \end{array}\right]\\
\label{LVM112}&=&y^m\left\{\begin{array}{rl}&\Big[\displaystyle-i\frac{d}{d\tau}+m(\al^\prime k^2+2)\Big]V_{M_1}^{\mu}\\
+&im^2V_{P_1}^\mu+\displaystyle\frac{1}{3}(2m^3-3m^2+m)i\al^\prime k^\mu V_T\end{array}\right\}.
\eeqa
Finally, the conformal algebra for the massive spin-two vertex operator $\tilde V_M$ can be obtained from the weighted sum of Eq.\eqref{LVM12} and Eq.\eqref{LVM112},
\beqa
\big[L_m,\tilde V_M\big]
\notag&=&\displaystyle\frac{\epsilon_{\mu\nu}}{2\al^\prime}\big[L_m, V_{M_1}^{\mu\nu}\big]-i\frac{\epsilon_\mu}{\sqrt{2\al^\prime}}\big[L_m, V_{M_1}^\mu\big] \\
\label{LVMM}&=&y^m\left\{\begin{array}{rl}\Big[-&\displaystyle i\frac{d}{d\tau}+m(\al^\prime k^2+2)\Big]\tilde V_M\\
+&\displaystyle\frac{m^3}{6}\big(\epsilon_{\mu\nu}\eta^{\mu\nu}+2\sqrt{2\al^\prime}\epsilon\cdot k\big)V_T\\
+&\displaystyle m^2\Big(\epsilon_{\mu\nu}k^\nu+\frac{\epsilon_\mu}{\sqrt{2\al^\prime}}\Big)V_{P_1}^\mu-m^2\sqrt{\frac{\al^\prime}{2}}\epsilon\cdot kV_T\\
-&\displaystyle m \epsilon_{\mu\nu}k^\nu V_{P_1}^\mu-\frac{m}{6}\big(\epsilon_{\mu\nu}\eta^{\mu\nu}-2\sqrt{2\al^\prime}\epsilon\cdot k\big) V_T\end{array}\right\}.
\eeqa
\subsection{Computation of the Conformal Algebra $\big[L_m,V_{M}\big]$}
\par
To complete the calculation of the conformal algebra for massive spin-two vertex operator $V_M$, we need two contributions from $V_{P_1}^\mu$ and $V_T$ in Eq.\eqref{VM},
\beqa
\label{LVM2}\epsilon_{\mu\nu}k^\nu\big[L_m,V_{P_1}^\mu\big]&=&
y^m\left\{\begin{array}{rl}\Big[-&\displaystyle i\frac{d}{d\tau}+m(\al^\prime k^2+2)\Big]\epsilon_{\mu\nu}k^\nu V_{P_1}^\mu\\
+&\displaystyle m^2\al^\prime\epsilon_{\mu\nu}k^\mu k^\nu V_T\\
-&\displaystyle m\big(\epsilon_{\mu\nu}k^\mu V_{P_1}^\mu+\al^\prime\epsilon_{\mu\nu}k^\mu k^\nu V_T\big)\end{array}\right\},
\eeqa
\beqa
\label{LVM3}\Om\big[L_m,V_T\big]&=&\Om y^m\left\{\begin{array}{rl} \Big[-&\displaystyle i\frac{d}{d\tau}+m(\al^\prime k^2+2)\Big]V_T\\
-&2m V_T\end{array}\right\}.
\eeqa
Notice that we have adjusted the conformal weight dimensions in both equations and made compensation by subtracting the extra terms. Putting everything together, Eqs.\eqref{LVMM}, \eqref{LVM2}, \eqref{LVM3}, we derive the complete conformal algebra for $V_M$ as follows:
\beqa
\notag\big[L_m,V_M\big]&=&
~~~y^m\Big[-i\frac{d}{d\tau}+m(\al^\prime k^2+2)\Big]V_M\\
\label{LLVM}&&+y^m
\left[\begin{array}{rl}
   \displaystyle\frac{m^3}{6}& \displaystyle\Big(\epsilon_{\mu\nu}\eta^{\mu\nu}+2\sqrt{2\al^\prime}\epsilon\cdot k\Big)V_T \\
  +m^2 & \displaystyle\Big(\epsilon_{\mu\nu}k^\nu+\frac{\epsilon_\mu}{\sqrt{2\al^\prime}}\Big) V_{P_1}^\mu \\
  -m^2 & \displaystyle\Big(\al^\prime\epsilon_{\mu\nu}k^\mu k^\nu+\sqrt{\frac{\al^\prime}{2}}\epsilon\cdot k\Big)V_T \\
 \displaystyle -\frac{m}{6} &  \displaystyle\Big(-6\al^\prime\epsilon_{\mu\nu}k^\mu k^\nu+\epsilon_{\mu\nu}\eta^{\mu\nu}-2\sqrt{2\al^\prime}\epsilon\cdot k+12\Om\Big)
\end{array}\right].
\eeqa
By comparing Eq.\eqref{LLVM} with the canonical formula, Eq.\eqref{ComforAlg}, we observe that: (1) The first line of the R.H.S. of Eq.\eqref{LLVM} gives the on-shell condition of the massive spin-two state $\al^\prime k^2=-1$. (2) The $m^3$ terms of the R.H.S. of Eq.\eqref{LLVM} give the $L_2$ condition for polarization tensors ($\epsilon_{\mu\nu}$,$\epsilon_\mu$). (3) The $m^2$ terms of the R.H.S of Eq.\eqref{LLVM} give the $L_1$ condition for polarization tensors ($\epsilon_{\mu\nu}$,$\epsilon_\mu$). (4) The coefficient of $m$ terms vanishes identically.
\section{Summary}\label{VI}
\par In this paper, we study the conformal algebra associated with the bosonic open string vertex operators in flat space-time. It is shown that normal-ordering of the covariant vertex operators leads to descending formulae, Eqs.\eqref{VP} and \eqref{VM}, and we can perform the computations of the conformal algebra in a recursive manner. At the massless level, the descending formula does not give preference to photon vertex operators with different normal-ordering prescriptions, i.e. $V_P$ v.s. $V_{P_1}$. However, in the case of massive vertex operators, fully covariant Virasoro conditions pick up a special normal-ordering. While $\tilde V_M$ vertex operator, Eq.\eqref{TVM}, satisfies the conformal algebra, Eq.\eqref{LVMM}, only with the complete gauge-fixed polarizations ($\epsilon_{\mu}=0=\epsilon_{\mu\nu}k^\nu$, $\epsilon_{\mu\nu}\eta^{\mu\nu}=0$), it is $V_M$ vertex operator which is compatible with the fully covariant Virasoro constraints, Eqs.\eqref{L0M}, \eqref{L1M}, \eqref{L2M}. From our prescription, it is easy to generalize to higher massive levels and one can show that these massive vertex operators satisfy the corresponding fully covariant Virasoro constraints.
\appendix
\section{Useful Formulae in The Calculations of Conformal Algebra}\label{A}
For normal-ordering calculation,
\beqa
\notag\Big[\dot X^\mu_+, \exp\big(ik\cdot X_{-}\big)\Big]&=&\Big[\sqrt{2\al^\prime}\sum_{n=1}^\infty\al_{n}^\mu e^{-in\tau},\exp\Big(\sqrt{2\al^\prime}k\cdot\sum_{m=1}^{\infty}\frac{\al_{-m}}{m}e^{im\tau}\Big)\Big]\\
\label{A-1}&=&2\al^\prime k^\mu\exp\big(ik\cdot X_{-}\big)\sum_{n=1}^{\infty}\mathbf{1}=-\al^\prime k^\mu\exp\big(ik\cdot X_{-}\big).\\
\notag\Big[\dot X^\mu_+, \dot X^\nu_-\Big]&=&
\Big[\sqrt{2\al^\prime}\sum_{n=1}^{\infty}\al_n^\mu e^{-in\tau},\sqrt{2\al^\prime}\sum_{n=1}^{\infty}\al_{-m}^\nu e^{im\tau}\Big]\\
 \label{A-2}&=&2\al^\prime\eta^{\mu\nu}\Big(\sum_{n,m=1}^{\infty}n\Big)=-\frac{\al^\prime}{6}\eta^{\mu\nu}.\\
\notag \Big[\ddot X^\mu_+, \exp\big(ik\cdot X_{-}\big)\Big]&=&\Big[ -i\sqrt{2\al^\prime}\sum_{n=1}^\infty n\al_{n}^\mu e^{-in\tau},\exp\big(\sqrt{2\al^\prime}k\cdot\sum_{m=1}^{\infty}
 \frac{\al_{-m}}{m}e^{im\tau}\big)\Big]\\
 \label{A-3}&=&-2i\al^\prime k^\mu\exp\big(ik\cdot X_{-}\big)\Big(\sum_{n=1}^{\infty}n\Big)=\frac{ik^\mu\al^\prime}{6}\exp\big(ik\cdot X_{-}\big).
\eeqa
Here we collect all relevant formulae in the calculations of conformal algebra:
\beqa
\label{A1}\frac{d}{d\tau}e^{ik\cdot X_0}&=&\big(ik\cdot \dot X_0-i\al^\prime k^2\big)e^{ik\cdot X_0}=e^{ik\cdot  X_0}\big(ik\cdot\dot X_0+i\al^\prime k^2\big).\\
\label{A2}\frac{d}{d\tau}e^{ik\cdot Z}&=&\frac{d}{d\tau}\big(e^{ik\cdot X_0}e^{ik\cdot X_+}\big)=\big(ik\cdot\dot Z-i\al^\prime k^2\big)e^{ik\cdot Z}.\\
\label{A3}\big[L_m,e^{ik\cdot X_-}\big]&=&y^me^{ik\cdot X_-}\Big[\sqrt{2\al^\prime}\sum^{m-1}_{n=1}(k\cdot\al_n)y^{-n}+k\cdot\dot Y+\al^\prime k^2(m-1)\Big],\\
\label{A4}\big[L_m,e^{ik\cdot Z}\big]&=&y^m\Big[\sqrt{2\al^\prime}\sum^{\infty}_{n=m}(k\cdot\al_n)y^{-n}\Big]e^{ik\cdot Z}.\\
\label{A5}\big[L_m,W^{\mu\nu}\big]&=&y^m\Big[-i\frac{d}{d\tau}W^{\mu\nu}+2mW^{\mu\nu}+\frac{\al^\prime}{3}m(m^2-1)\eta^{\mu\nu}\Big].\\
\label{A6}\big[k\cdot \al_n,W^{\mu\nu}\big]&=&\sqrt{2\al^\prime}ny^n\big(k^\mu\dot X^\nu+k^\nu\dot X^\mu\big).
\eeqa
Here $W^{\mu\nu}\equiv \dot Y^\mu\dot X^\nu+\dot X^\nu\dot X_+^\mu$.
\begin{acknowledgments}
The authors wish to dedicate this paper to the string focus group in Taiwan. We appreciate all the education and inspiration we have received from all members and visitors of the group since its incarnation in 1998. This work is supported by the
 National Science Council of Taiwan under the contract 96-2112-M-029-002-MY3 and the string focus group under the  National Center for Theoretical Sciences.
\end{acknowledgments}

{}
\end{document}